\newcommand{\R}{\mathbb R}
\newcommand{\M}{\mathcal M}

\def\ddt{\frac{\text d}{\text dt}}

\documentclass[12pt,oneside]{amsart}

\usepackage{amsmath}               
\usepackage{times}
\usepackage{hyperref}

\numberwithin{equation}{section}

\title[Gravitational collapse of homogeneous scalar fields]%
{Gravitational collapse of homogeneous scalar fields}

\author[R.\ Giamb\`o]{Roberto Giamb\`o}
\address{Dipartimento di Matematica e Informatica,
Universit\`a di Camerino, Italy} \email{roberto.giambo@unicam.it}
\urladdr{http://dmi.unicam.it/\~{}giambo}
\date{April 2005 (revised version)}

\theoremstyle{plain}\newtheorem{teo}{Theorem}[section]
\theoremstyle{plain}\newtheorem{prop}[teo]{Proposition}
\theoremstyle{plain}
\theoremstyle{plain}
\theoremstyle{definition}
\theoremstyle{remark}\newtheorem{rem}[teo]{Remark}
\theoremstyle{definition}\newtheorem{example}[teo]{Example}

\theoremstyle{plain}

\begin{document}

\begin{abstract}
Conditions under which gravity coupled to self interacting scalar
field determines singularity formation are found and discussed. It
is shown that, under a suitable matching with an external space, the
boundary, if  collapses completely, may give rise to a naked
singularity. Issues related to the strength of the singularity are
discussed.
\end{abstract}

\maketitle

\section{Introduction}\label{sec:intro}
The gravitational collapse of scalar fields in classical General
Relativity has been widely studied in literature. Its role in
understanding the machinery ruling singularities' causal structure
-- at least in spherical symmetry -- was clear since 90's, when
models of scalar field collapse exhibiting a naked singularity, and
therefore violating the so--called Penrose's Cosmic Censorship
Conjecture \cite{pen}, were found numerically by Choptuik
\cite{chop} and analytically by Christodoulou \cite{chr1}.

In these pioneering works the scalar field is massless and free,
that means that the Lagrangian function of the field does not
contain any mass or potential terms. Recently, Hertog, Horowitz and
Maeda \cite{hor} found a class of potentials where smooth initial
data evolve to give rise to a naked singularity, although energy
conditions may be violated. Moreover, in a recent paper, Goswami and
Joshi \cite{joshi} studied collapse of self interacting scalar
field, under a homogeneity assumption, considering a class of models
where divergence of energy density near the singularity is assumed
to satisfy a power--law relation.

As is well known scalar field have not been observed although their fundamental relevance in cosmology.
Observational cosmology is indeed probably the unique field in which we can
hope to obtain information on the scalar field potential, and several works go in this direction,
as e.g. \cite{nunes} (and references therein) in which the analysis of transition from
matter domination to dark energy domination is taken into account, although the expanding case is
obviously considered there.

In the present paper, we consider homogeneous scalar fields collapsing models
\emph{in general}, characterizing the class of potentials
determining singularity formation. A crucial role in determining the
causal structure is played by the existence, or not, of apparent
horizon during the evolution. Since the singularity is (of course)
synchronous, the behavior is quite different from what happens in
many examples of matter models exhibiting a central naked
singularity \cite{JoshiCMP, ns}. Here, instead, it may be that the
singularity located at the boundary of the "ball" of scalar field
can be naked. Therefore, the scalar field solution must be matched
with a suitable external solution and behavior of radial geodesic in
the external solution must be studied accordingly.

Scalar field solution under homogeneity assumption is introduced in
Section \ref{sec:eqs}. In Section \ref{sec:sing}, necessary and
sufficient conditions for singularity formation are stated, together
with conditions which forbid apparent horizon formation during the
evolution. In the latter case, a matching with anisotropic
generalization of de Sitter spacetime \cite{aniso} is performed
(Section \ref{sec:deSitter}), and naked singularity existence is
proved -- and its gravitational strength discussed -- in Section
\ref{sec:naked}. Reasons for our choice of exterior region are
discussed in final Section \ref{sec:final}, together with overall
conclusions.

\section{The solution}\label{sec:eqs}
The general scalar field spacetime is a spacetime $(\M,g_{\mu\nu})$
satisfying Einstein field equations for the energy--momentum tensor
\begin{equation}\label{eq:Tgen}
T_{\mu\nu}=\partial_\mu\phi\partial_\nu\phi-\left(\frac12
g^{\alpha\beta}\partial_\alpha\phi\partial_\beta\phi+V(\phi)\right)g_{\mu\nu},
\end{equation}
where $\phi$ is a scalar function on $\M$ and $V$ is the potential.
We consider the case of a spherically symmetric spacetime where the
gradient of $\phi$ is timelike, which allows us to choose a comoving
gauge such that $\phi$ is a function of comoving time only, and the
metric and the energy--momentum tensor can be respectively written
in the form \cite{kra}
\begin{equation}\label{eq:metric}
\text ds^2=-\text dt^2+a^2(t)\,\text dr^2+ a^2(t) r^2\,\text
d\Omega^2,
\end{equation}
and
\begin{equation}\label{eq:T}
T^\mu_\nu=\text{diag}(-\epsilon,p,p,p)
\end{equation}
Equations \eqref{eq:Tgen}--\eqref{eq:metric} imply that, in the
above expression, the energy density $\epsilon$ and the stress $p$
can be written in terms of the field and the potential as
\begin{equation}\label{eq:ep}
\epsilon=\frac12\dot\phi^2+V(\phi),\qquad
p=\frac12\dot\phi^2-V(\phi),
\end{equation}
where the dot denotes differentiation with respect to $t$.

Field equations reduce to
\begin{align}
& \dot a^2=\frac{8\pi}3 a^2\epsilon,\label{eq:dota}\\
& \ddt\left(a^3\dot\phi\right)=-a^3 V'(\phi).\label{eq:KG}
\end{align}
in the unknown functions $a(t)\ge 0$ and $\phi(t)$. The potential
$V(\phi)$ has to be considered as a constitutive function that,
analogously to what happen in the elastic--solid matter case
\cite{magli},  establishes  a relation between energy and stresses.
In particular, \eqref{eq:KG} is Klein--Gordon equation
$g^{\mu\nu}\nabla_\mu\partial_\nu\phi=V'(\phi)$, obviously
equivalent to the Bianchi identity .

We will take initial data at an initial time, say $t=0$:
\begin{equation}\label{eq:initial}
a(0)=a_0,\quad\phi(0)=\phi_0,\quad\dot\phi(0)=\dot\phi_0.
\end{equation}
The field evolves collapsing until it possibly reaches the singular
state at $a=0$. We will consequently consider cases where $a$ is a
monotonically decreasing function of $t$. This fact implies that,
for instance, the energy density $\epsilon$ can be seen as a
function of $a$, and can be written as
\begin{equation}\label{eq:eps}
\epsilon=\frac{3}{8\pi}\left(\frac{\psi(a)}a\right)^2,
\end{equation}
where $\psi\in C^\infty((0,a_0),(0,+\infty))$. Note that, in
principle, we don't make assumption on the behavior of $\psi(a)$ as
$a\to 0^+$ but, in order to study a physically meaningful
singularity, we will assume that
\begin{equation}\label{eq:assum}
\lim_{a\to 0^+}\frac{\psi(a)}a=+\infty.
\end{equation}
Using \eqref{eq:eps}, equation \eqref{eq:dota} takes the form
\begin{equation}\label{eq:dota2}
\dot a=-\psi(a),
\end{equation}
where the sign choice is determined by the positive sign of
$\psi(a)$ and the fact that we are considering a collapsing
situation. Therefore, using initial condition \eqref{eq:initial},
the following proposition easily follows.
\begin{prop}\label{thm:a}
The metric \eqref{eq:metric} is determined by the function
\begin{equation}\label{eq:a(t)}
a(t)=F^{-1}(F(a_0)-t),
\end{equation}
where $F=F(a)$ is a primitive function of $\,1/\psi(a)$.
\end{prop}
Moreover, such a choice of $\psi$ also determines both the scalar
function $\phi$ and the potential $V$ as functions of $a$. Indeed,
with the positions
\begin{equation}\label{eq:Rm}
R:=a(t) r,\qquad m:=\frac R2\left(1-g^{\alpha\beta}\partial_\alpha
R\,\partial_\beta R\right)=\frac12 r^3 a(t) \dot a^2(t),
\end{equation}
using \eqref{eq:ep}, \eqref{eq:dota2}, it is easy to see that
Klein--Gordon equation \eqref{eq:KG} is completely equivalent to
\begin{equation}\label{eq:dotm}
\dot m=-4\pi\,p\, R^2\,\dot R,
\end{equation}
and, in view of the above equation and \eqref{eq:dota2} again, we
can write the stress $p$  as
\begin{equation}\label{eq:efe-p}
p=-\frac{\psi^2(a)}{8\pi
a^2}\left(1+2a\frac{\psi'(a)}{\psi(a)}\right).
\end{equation}
Using \eqref{eq:ep}, \eqref{eq:dota2}, one finally finds
\begin{equation}\label{eq:phi}
\left(\frac{\text d\phi}{\text da}\right)^2=\frac1{4\pi
a^2}\left(1-a\frac{\psi'(a)}{\psi(a)}\right),
\end{equation}
\begin{equation}\label{eq:V}
V=\frac{\psi^2(a)}{4\pi
a^2}\left(1+\frac12a\frac{\psi'(a)}{\psi(a)}\right).
\end{equation}
Therefore, chosen an energy profile through the function $\psi(a)$,
the solution is completely expressed in function of $a$ and,
consequently, of $t$. In case the function $\phi(a)$ is invertible,
one can say more. Indeed, the following proposition holds.
\begin{prop}\label{thm:Vphi}
If $\tfrac{\psi'(a)}{\psi(a)}a<1$ in $(0,a_0)$, the spacetime given
by \eqref{eq:a(t)} solves Einstein field equations, with the
potential $V(\phi)=V(a(\phi))$, where $V(a)$ is given by
\eqref{eq:V}, and $a(\phi)$ is the inverse function of
\begin{equation}\label{eq:phi2}
\phi(a)=\phi_0+\int_a^{a_0}\sqrt{\frac1{4\pi
\theta^2}\left(1-\theta\frac{\psi'(\theta)}{\psi(\theta)}\right)}\,\text
d\theta.
\end{equation}
\end{prop}
\begin{rem}\label{rem:wec}
Since
$$\epsilon+p=\dot\phi^2=\dot a^2\left(\frac{\text d\phi}{\text
da}\right)^2,$$ the condition $\tfrac{\psi'(a)}{\psi(a)}a<1$ in
$(0,a_0)$ can be reinterpreted by saying that the weak energy
condition inequality must be \emph{strictly} satisfied, that is
$\epsilon+p>0$. Of course, one can invert the function $\phi(a)$
\eqref{eq:phi2} under more general conditions also, for instance if
$\epsilon+p$ vanishes for some isolate values of $a$, but in this
case problems of regularity may arise.
\end{rem}

\section{Singularity formation}\label{sec:sing}
Properties of $\psi(a)$ will be used to study singularity formation
and behavior for the solutions above. The following theorem holds.
\begin{teo}\label{thm:form}
The spacetime becomes singular in a finite amount of comoving time
if and only if the function $1/\psi(a)$ is integrable on $(0,a_0)$.
The time of collapse is given by
\begin{equation}\label{eq:ts}
t_s=\int_0^{a_0}\frac{1}{\psi(a)}\,\text da.
\end{equation}
\end{teo}
\begin{proof}
If $1/\psi(a)$ is integrable in $(0,a_0)$, the primitive $F(a)$ can
be continuously extended to $a=0$. In this case $t_s=F(a_0)-F(0)$,
and using \eqref{eq:a(t)} it follows that $a(t_s)=0$. Viceversa, if
$a(t)=0$ for some $t_s\in(0,+\infty)$, then by continuity
$\lim_{t\to t_s} a(t)=0$. Equation \eqref{eq:a(t)} again implies
that $\lim_{a\to 0^+}F(a)=F(a_0)-t_s$, and the primitive can be
extended up to $a=0$.
\end{proof}

We will suppose hereafter that we are in the hypothesis of Theorem \ref{thm:form},
so that singularity forms in a finite amount of comoving time.
In order to study singularity behavior, we fix $t_s$ as in
\eqref{eq:ts} and consider the \emph{apparent horizon} curve
$t_h(r)$ implicitly defined by the equation
$R(r,t_h(r))=2m(r,t_h(r))$. It is the boundary of the \emph{trapped
region}
\[
\mathcal T=\{(r,t)\,:\,R(r,t)\le 2m(r,t)\}.
\]

The following result holds.
\begin{prop}\label{thm:nohor}
If $\psi(a)$ is bounded in $(0,a_0)$, there exists $r_b>0$ such
that, for any shell of matter $r\leq r_b$, no apparent horizon forms
during the evolution.
\end{prop}

\begin{proof}
Using \eqref{eq:Rm}, the inequality $R>2m$ means that
\begin{equation}\label{eq:ext}
\psi(a) r<1.
\end{equation}
Then, it suffices to choose $r_b<(\sup_{(0,a_0)}\psi(a))^{-1}$, and,
if $r\le r_b$ and $t<t_s$,  it will be $(t,r)\not\in\mathcal T$.
\end{proof}

\begin{rem}\label{rem:bound}
The quantity $r_b$ will play the role of the boundary of scalar
field collapsing sphere. Note that the absence of horizon does not
mean, in principle, that the singularity is naked. Indeed, if a
future pointing null geodesic arising from the singularity exists,
it must satisfy the ODE
\begin{equation}\label{eq:ODE}
\frac{\text dt}{\text dr}=a(t),
\end{equation}
and then it must be an increasing function of $r$. This cannot
happen for $r\in[0,r_b)$, because the singularity is synchronous.
But it could be the case for a null geodesic to arise from
$(r_b,t_s)$ and go into the external space. This situation can be
studied only after we join, at the shell $r_b$, the solutions found
in Section \ref{sec:eqs} with an exterior solution. This will be
done in Section \ref{sec:deSitter}.
\end{rem}

\begin{rem}\label{rem:counterex}

Note that one can conceive cases of such no trapped surface
formation when pressures \eqref{eq:efe-p} \emph{do not diverge} and
\emph{do not remain negative} in the approach to the singularity,
even when w.e.c. is satisfied.  Take, for instance, $
\psi(a)=\sqrt[5]{a}(\sin\log a+\tfrac53). $ Anyway, if pressures
diverge, potential also (positively) does.
\end{rem}

We now consider the case where $\psi$ is unbounded, positively
diverging at the singularity.

\begin{prop}\label{thm:hor}
If $\lim_{a\to 0^+}\psi(a)=+\infty$, for any $r>0$ such that the
initial data are taken outside the trapped region $\mathcal T$, the
shell labelled $r$ becomes trapped strictly before it becomes
singular, and so a black hole forms.
\end{prop}
\begin{proof}
Fix any $r>0$. If $(r,0)\not\in\mathcal T$, then
$\psi(a_0)<\tfrac1{r}$. But $\psi(a)\to+\infty$ as $a\to 0$, and
therefore by continuity there must exists a time $t_h(r)<t_s$ such
that $\psi(a(t_h(r)))=1/r$. Then, the apparent horizon lies below
the singularity curve $t=t_s$, which is therefore covered for any
$r>0$.

Similar arguments to those made in Remark \ref{rem:bound} show that,
even if $\lim_{r\to 0^+}t_h(r)=t_s$, there cannot exist future
pointing null geodesic below the horizon arising from the centre,
and then $t=t_s$ is covered at each shell.
\end{proof}

\begin{rem}\label{rem:open}
Note that hypothesis of Proposition \ref{thm:hor} does not cover all
cases of unbounded $\psi(a)$, which indeed may not have limit as
$a\to 0^+$.
\end{rem}

\begin{example}\label{ex:power}
Consider the case of $\psi$ ruled by the power--law relation
\cite{joshi}
$$
\psi(a)=\sqrt{\frac{8\pi}{3}}\, a^\beta,\quad\beta\in\R,
$$
where $\beta<1$ by compatibility with \eqref{eq:assum}. Hypotheses
of Theorem \ref{thm:form} are always verified with this choice of
$\psi$, therefore the model collapses in a finite amount of comoving
time. Using the above equations one finds
\begin{align*}
&a(t)=\left(a_0^{1-\beta}-\sqrt{\frac{8\pi}{3}} (1-\beta)
t\right)^{\frac1{1-\beta}},\\
&\phi(a)=\phi_0+\sqrt{\frac{1-\beta}{4\pi}}\log\frac{a_0}a,\\
&V(\phi)=V_0 e^{\sqrt{16\pi(1-\beta)}(\phi-\phi_0)}.
\end{align*}
If $\beta$ is not negative we are in the case of Proposition
\ref{thm:nohor} and no apparent horizon forms if one takes $r_b$
sufficiently small. On the other side, $\beta$ negative implies
apparent horizon existence (Proposition \ref{thm:hor}), and
$$
t_h(r)=t_s-\gamma r^{\frac{\beta-1}\beta},
$$
where $\gamma>0$ depends on $\beta$. Note that the centre gets
trapped at singular comoving time $t_s$. The particular case
$\beta=-\tfrac12$ corresponds -- as easily inferred using
\eqref{eq:ep} -- to vanishing stresses, and in fact the solution is
formally equivalent to the case of homogeneous dust cloud collapse
(Oppenheimer--Snyder model \cite{opp}).

\end{example}

\section{The exterior region}\label{sec:deSitter}
Our aim is to join the scalar field spacetime with an exterior
solution at $r=r_b$, and study singularity arising from
$r=r_b,\,t=t_s$. The matching will be performed along a hypersurface
$\Sigma$ using Israel--Darmois junction conditions \cite{darm, isr},
that requires continuity across $\Sigma$ of the first and the second
fundamental forms induced on $\Sigma$ by the two spacetimes. For
this reason, one cannot expect to perform the junction with
Schwarzschild line element, since radial stresses of the scalar
field ball do not vanish in principle (see \eqref{eq:efe-p}).
Therefore, as exterior solution, we will consider the anisotropic
generalization of de Sitter spacetime -- see \cite{aniso} and
references therein -- that we here briefly review.

Consider a spherically symmetric spacetime such that the metric
tensor and the energy--momentum tensor respectively reads
\begin{align}
&\text ds^2=-e^{2\nu}\text dt^2 +e^{2\lambda}\text dr^2 +Y^2 (\text
d\theta^2
+\sin^2\theta\, \text d\phi^2),\label{eq:ds}\\
&T^\mu_\rho={\rm
diag}\left(-\epsilon(r,t),p_r(r,t),p_t(r,t),p_t(r,t)\right).\label{eq:Tss}
\end{align}
As well known, stresses are isotropic in case of de Sitter solution
$p_r=p_t=p$, where $\epsilon+p=0$. One may try to look for
generalizations of this case, braking the isotropy conditions, that
is supposing $p_r\ne p_t$, but still assuming
\begin{equation}\label{eq:deSitter}
\epsilon+p_r=0.
\end{equation}
This amounts to weaken the degeneracy hypothesis on the tensor
$T^\mu_\nu$ of de Sitter solution, retaining two different
degenerate subspaces. Misner--Sharp mass  reads
\begin{equation}\label{eq:M-S}
m={\frac Y2}\left[1-(Y'e^{-\lambda})^2+(\dot Y e^{-\nu})^2\right],
\end{equation}
(in this Section, dot and prime will denote differentiation w.r.t $t$ and
$r$, respectively). Using Einstein equations together with
assumption \eqref{eq:deSitter} one shows that both $\epsilon$ and
$m$ can be seen as functions of $Y$. In particular
$m(Y)=4\pi\int_0^Y\epsilon(\sigma)\sigma^2\,\text d\sigma+m_0$. With the
additional ansatz that
$$
(Y'e^{-\lambda}),\quad (\dot Y e^{-\nu}),
$$
are functions of the variable $Y$ only,
that amounts to require a higher degree of symmetry of the solution
-- a $G_4$ group of motion -- it can be seen that a suitable
coordinate change exists, that brings the solution in the form
\begin{equation}\label{eq:KS}
\text ds^2=-\chi(Y)\,\text dT^2+\chi(Y)^{-1}\,\text dY^2+Y^2\,\text
d\Omega^2,\quad\chi(Y)=1-\frac{2M(Y)}Y,
\end{equation}
thereby obtaining a family of solution as the  mass $M(Y)$ varies.
This function $M(Y)$ is arbitrary but, to satisfy weak energy
conditions, it must obey to the following law
\begin{equation}\label{eq:wec2}
M'(Y)\ge 0,\qquad M''(Y)-\frac 2Y M'(Y)\le 0.
\end{equation}
The family outlined here contains, for instance, Minkowski,
Schwarzschild, Reissner--Nordstr\"om and de Sitter spacetimes as
particular cases, corresponding to choosing the function $M(Y)$
respectively equal to 0, to $m_0$ (constant), to $m_0+e_0^2/Y$
($m_0,\,e_0$ constant) and to $\frac 43\epsilon_0 \pi Y^3$.
Computation of Kretschmann scalar implies that the only singularity
may arise at $Y=0$ -- for instance, it suffices that
$m(Y)=Y^\alpha+o(Y^\alpha)$ for $Y\to 0$, with $0<\alpha<3$.

\subsection{Matching conditions}\label{subsec:match}
Let us now consider a spherical symmetric source
\begin{equation}\label{eq:dsint}
\text ds^2=-e^{2\nu}\text dt^2 +e^{2\lambda}\text dr^2 +R^2(r,t)
(\text d\theta^2 +\sin^2\theta\, \text d\phi^2),
\end{equation}
Our aim is to show that, at a spherical junction hypersurface
$r=r_b$, no conditions other than continuity of the mass function
are required to perform the matching. Indeed, the following
proposition holds.
\begin{prop}\label{thm:join}
A general spherical line element \eqref{eq:dsint} can be joined with
the spacetime \eqref{eq:KS} at a hypersurface
$\Sigma:=\{(t,r=r_b,\theta,\phi)\}$. The matching conditions at
$\Sigma$ read
\begin{equation}\label{eq:cond}
Y(t)=R(r_b,t),\qquad\frac{\text d T}{\text d
t}(t)=\frac{R'(r_b,t)}{\chi(m(r_b,t))e^{(\lambda-\nu)(r_b,t)}},
\end{equation}
where $m(r,t)$ is the Misner--Sharp mass function of the metric
\eqref{eq:dsint}.
\end{prop}

\begin{proof}
Let us parameterize $\Sigma$ with coordinates $(\tau,\theta,\phi)$.
The injection of $\Sigma$ into the internal space equipped with
metric \eqref{eq:dsint} simply reads
$(\tau,\theta,\phi)\hookrightarrow(\tau,r_b,\theta,\phi)$.
Consequently, the fundamental forms induced by the metric on
$\Sigma$ can be computed as
\begin{align}
& I^{int}_\Sigma= -e^{2\nu}\,\text d\tau^2+R^2\,\text
d\Omega^2,\label{eq:int1}\\
&I\!\!I^{int}_\Sigma= -e^\lambda\left[e^{2(\nu-\lambda)}\nu'\,\text
d\tau^2-e^{-2\lambda}R'R\,\text d\Omega^2\right].\label{eq:int2}
\end{align}
where we suppose the metric terms calculated in $(r_b,t)$.

We perform the same operation with the external metric
\eqref{eq:KS}. The injection reads in coordinates
$(\tau,\theta,\phi)\hookrightarrow(T(\tau),Y(\tau),\theta,\phi)$,
where $T(\tau)$, $Y(\tau)$ are unknown functions. First fundamental
form in this case is
\begin{equation}\label{eq:ext1}
I^{ext}_\Sigma=\left(-\chi(Y)\dot T^2+\frac{\dot
Y^2}{\chi(Y)}\right)\,\text d\tau^2+ Y^2\,\text d\Omega^2,
\end{equation}
where, with a slight abuse of notation, we denote derivatives of $T$
and $Y$ by $\dot T$ and $\dot Y$. Continuity of the first
fundamental form at $\Sigma$ implies
\begin{align}
&Y(\tau)=R(\tau,r_b),\label{eq:1}\\
\intertext{which is the first equation in \eqref{eq:cond}, and}
&-\chi(Y)\dot T^2+\frac{\dot Y^2}{\chi(Y)}=-e^{2\nu}.\label{eq:2}
\end{align}
The second fundamental form reads ($M_{,Y}$ is the derivative of
$M(Y)$)
\begin{multline}\label{eq:ext2}
I\!\!I^{ext}_\Sigma= e^{-\nu}\left\{\dot Y\left[\ddot
T+2\left(\frac{M}{Y^2}-\frac{M_{,Y}}{Y}\right)\frac{\dot Y\dot
T}{\chi(Y)}\right]\,\text d\tau^2-\right.\\
\left.-\dot T\left[\ddot
Y+\left(\frac{M}{R^2}-\frac{M_{,Y}}{Y}\right)e^{2\nu}\,\text
d\tau^2-Y\chi(Y)\,\text d\Omega^2\right]\right\},
\end{multline}
where we have used  \eqref{eq:2}. Comparison of the angular terms in
the two second fundamental forms immediately yields
\begin{equation}\label{eq:cond2}
\dot T=\frac{R'}{\chi e^{(\lambda-\nu)}}
\end{equation}
i.e. the second condition in \eqref{eq:cond}. Finally, let us show
that there are no other conditions to be imposed. First, substituting
\eqref{eq:cond} in \eqref{eq:2} we get the identity
\begin{equation}\label{eq:mass}
\chi(R)=(R' e^{-\lambda})^2-(\dot R e^{-\nu})^2,
\end{equation}
that only says that the mass function must be continuous across
$\Sigma$, that is $m(r_b,t)=M(Y(t))$. Therefore, we are only left
with the proof that coefficients of $\text d\tau^2$ in the two second
fundamental forms coincide, that is we have to prove
\begin{equation}\label{eq:toprove1}
e^{-\nu}\left[\ddot T\dot Y-\dot T\ddot
Y\left(\frac{M}{Y^2}-\frac{M_{,Y}}{Y}\right)\left(\frac2\chi\dot
Y^2\dot T-e^{2\nu}\dot T\right)\right]+e^{2\nu-\lambda}\nu'=0
\end{equation}
But this equation can be easily shown to hold identically by
calculating the quantities $\ddot T$ from \eqref{eq:cond2} and
$\ddot Y$ from \eqref{eq:mass}, and using Einstein equation $\dot
R'-\dot\lambda R'-\nu'\dot R=0$.
\end{proof}

\begin{rem}\label{rem:Vai}
The anisotropic generalizations of de Sitter spacetime \eqref{eq:KS}
are formally a subclass of the so--called \emph{generalized Vaidya
solutions} \cite{gov,wang}, given by
\begin{equation}\label{eq:Va}
\text ds^2=-\left(1-\frac{2M(V,Y)}Y\right)\,\mathrm dV^2-2\,\mathrm
dY\,\mathrm dV + Y^2\,\mathrm d\Omega^2.
\end{equation}
Indeed, in case $M$ depends only on $Y$, i.e. $M_{,V}=0$, the
coordinate transformation $\mathrm dV=\mathrm
dT-\tfrac1{\chi(Y)}\mathrm dY$ brings \eqref{eq:KS} into
\eqref{eq:Va}.

Of course, a similar result to Proposition \ref{thm:join} can be
proved for this wider class of solutions. But, in this case, there
is an additional condition on the mass function, that must satisfy
the requirement (see also \cite[equation (39)]{joshi})
\begin{equation}\label{eq:addit}
M_{,V}=0\quad\text{on\ }\Sigma,
\end{equation}
to perform the matching with the general spherical symmetric line
element \eqref{eq:dsint}.
\end{rem}

\section{Naked singularities existence}\label{sec:naked}
We now confine ourselves to the case when $\psi(a)$ is a bounded
function of $a$.
The singularity that develops is therefore massless, as one can easily infer
from \eqref{eq:Rm} using \eqref{eq:dota2}. Moreover,
in view of Proposition \ref{thm:nohor}, the singular
curve $t=t_s$ does not get trapped if one take $r_b$ sufficiently
small.

In view of the results of Section \ref{sec:deSitter} above, we
consider the spacetime where the interior scalar field is matched
with an exterior anisotropic generalization of de Sitter spacetime
\eqref{eq:KS} at the shell $\Sigma$ labeled $r_b$. Let us recall
that the mass \eqref{eq:Rm} $m$ is equal, using \eqref{eq:dota2}, to
$\frac12 r^3 a\psi^2(a)=\frac12 r^2 R \psi^2(\tfrac Rr)$, and
therefore a necessary condition must be
\begin{equation}\label{eq:massext}
M(Y)=\frac12 r_b^2 Y\psi^2(\tfrac Y{r_b}),\qquad\forall Y\in[0,a_0^2 r_b^2].
\end{equation}

In principle, the matching condition fixes the mass function for the external
region for the portion of junction surface corresponding to the observed
scalar field collapsing ball, that is for $t$ running from 0 to the time of collapse
$t_s$. For bigger values of $Y$ ($Y>a_0^2 r_b^2$), the mass is determined by
the form of the metric for $t<0$.

As we already know, the exterior solution may possess a singularity
at $Y=0$, depending on the mass profile. From the equation of
state \eqref{eq:deSitter}, energy diverges at $Y=0$ if $p_r$ does,
and since, with the above choice of mass, radial pressure is
continuous across junction surface, then $Y=0$ is a singularity if
\begin{equation}\label{eq:divpress}
\lim_{a\to 0^+} p(a)=-\infty,
\end{equation}
where $p(a)$ is the quantity, depending on $\psi(a)$, given by \eqref{eq:efe-p}.

\begin{rem}\label{rem:wec}
Weak energy condition \eqref{eq:wec2} is equivalent, with the above
choice of $M(Y)$, to the following two conditions on the function
$\psi$:
\begin{equation}\label{eq:wec2new}
\psi(a)+2a\psi'(a)\ge
0,\quad\psi^2(a)-a^2\left(\psi'^2(a)+\psi(a)\psi''(a)\right)\ge 0.
\end{equation}
We observe by the way that the first of these inequalities implies
that potential \eqref{eq:V} diverges. Conditions \eqref{eq:wec2new}
are satisfied by power--law models of Example \ref{ex:power}, when
$\psi$ is bounded, i.e. for non-negative $\beta$.
\end{rem}

We prove the following theorem.

\begin{teo}\label{thm:naked}
Suppose that $\psi(a)$ is bounded and satisfies \eqref{eq:divpress},
and let $r_b$ such that $1-\psi^2(a(t))r^2$ is bounded away from 0
uniformly on $(0,r_b)\times(0,t_s)$. Then, the boundary
$\Sigma=\{r=r_b\}$ of the scalar field collapses to a naked
singularity.
\end{teo}

\begin{proof}
Equations \eqref{eq:cond} becomes, using \eqref{eq:dota2},
\eqref{eq:Rm} also,
\begin{equation}\label{eq:cond1}
Y(t)=a(t) r_b,\quad\frac{\text d T}{\text d
t}(t)=-\frac{r_b}{1-r_b^2\psi^2(a(t))}.
\end{equation}
Let $r_b$ such that $1-\psi^2(a(t))r^2$ is bounded away from 0. Such
a choice is, of course, always made possible  because $\psi(a)$ is
bounded. Then $\frac{r_b}{1-r_b^2\psi^2(a(t))}$ is bounded in
$(0,a_0)$ and we can integrate the ODE in \eqref{eq:cond1} to obtain
$$\lim_{t\to t_s} T(t)=T_0\in(0,+\infty).$$ Then the anisotropic
generalization of de Sitter spacetime solution can be extended up to
$T=T_0, Y=0$. But in a neighborhood of this point, $\chi$ is bounded
away from zero because junction conditions \eqref{eq:cond1} imply
continuity of the mass function, and by hypothesis,
$1-\frac{2m(t,r_b)}{Y(t,r_b)}$ is bounded away from zero as the
singularity is approached. Then the ODE
\begin{equation}
\frac{\text dT}{\text dY}=\frac{1}{\chi(Y)}
\end{equation}
can be solved in a right neighborhood of $Y=0$, and there exists a
radial null geodesic starting from the singularity, which is
therefore naked.

\end{proof}


\begin{rem}\label{rem:power1}
The above theorem applies to cases described in Example
\ref{ex:power}, with $\beta\ge 0$. We stress the fact that, in the
approach to the singularity $t=t_s$,
$\tfrac{2m}R=\psi(a(t))^2\,r_b^2$ does not need to vanish, but may
also tend to a constant nonzero value, or may even not possess a
limit value. Consider, for instance, Example \ref{ex:power} with
$\beta=0$. We know that that no horizon forms if $r_b$ is
sufficiently small, but now $1-\tfrac{2m}{R}$ is a positive constant
on $\Sigma$.
\end{rem}

\subsection{Strength of the singularity}\label{subsec:str}
Since we have found examples of spacetimes exhibiting a naked
singularity, it may be of interest to determine its \emph{strength}.
The first definition of strong curvature singularity was suggested
by Tipler \cite{tip} and studied e.g. in \cite{clarkrol,tipclark}.
Modifications to this definition have been also suggested by Nolan
\cite{nolan} and Ori \cite{ori} to take into account pathological
situations where, although volume forms along geodesics preserve,
Jacobi fields  have opposite irregular behavior -- i.e. one diverges
and another one vanishes in the approach to the singularity. For
most of our purposes, anyway, it will suffice to look at the
behavior of the quantity $k^2\Psi= k^2 R_{\alpha\beta} K^\alpha
K^\beta$ along causal geodesics, where $R_{\alpha\beta}$ is Ricci
tensor, $K^\alpha=\tfrac{\mathrm dx^\alpha}{\mathrm dk}$ is the
tangent vector of the geodesic with parameter $k$. If this quantity
remains bounded away from zero in the approach to the singularity,
this one may be considered as physically meaningful.

Let us first consider the metric \eqref{eq:metric}. If $K^\alpha$ is
a radial null geodesic, it can be computed, using \eqref{eq:dota2},
that
\[
k^2\Psi=2\left(\frac ka\frac{\mathrm da}{\mathrm
dk}\right)^2\left[1-\frac{\psi'(a)a}{\psi(a)}\right].
\]
If the limit $\tfrac{\mathrm da}{\mathrm dk}$ exists as $k\to 0$,
the quantity in round brackets tends to a finite value, while the quantity in
square brackets remains bounded away from zero if $\epsilon+p>0$
(see \eqref{rem:wec}), and then the radial null geodesic terminates
in a strong singularity.

Of course, in our model the interior scalar field region is matched
with an exterior where metric tensor is given by \eqref{eq:KS}, and
so strength of the central singularity of \eqref{eq:KS} must be
determined to see whether strong singularity persists.

\begin{prop}\label{thm:tip}
If,  $\forall a\in (0,a_0)$, the quantity
$\left(\frac{a\,\psi'(a)}{\psi(a)}+\frac{a\,\psi''(a)}{\psi'(a)}+2\right)$
is well defined and bounded away from zero, then there exists causal
geodesics terminating into the naked singularity and satisfying the
limit strong curvature condition $\liminf_{k\to 0}|k^2\Psi|>0$.
\end{prop}

\begin{proof}

Let us evaluate the strength of the $Y=0$ singularity in this case,
recalling that the mass $M(Y)$ is given by \eqref{eq:massext}.
Actually, it can now be checked  that $\Psi$ vanishes along null
geodesics, and so we will consider the situation for \emph{timelike}
radial geodesics. Their equations integrate to give (see also
\eqref{eq:cond1})
\[
\frac{\mathrm dT}{\mathrm dk}=\frac{\xi}{\chi(Y(k))},\qquad
\frac{\mathrm dY}{\mathrm dk}=\sqrt{\xi^2- \zeta^2\chi(Y(k))},
\]
($\xi$ and $\zeta\ne 0$ are integration constants) that is, they are
regular at $Y=0$. Therefore, considering a geodesic with
$\frac{\mathrm dY}{\mathrm dk}\to 0$, it can be seen that
\begin{multline*}
|k^2\Psi|=\frac{k^2}Y M''(Y)=\frac{2}{\mathrm d^2 Y/\mathrm dk^2}
M''(Y)=\\2\zeta^{-2}\frac{M''(Y) Y^2}{Y M'(Y)-M(Y)}=
2\zeta^{-2}\left(2+\frac{a\,\psi'(a)}{\psi(a)}+\frac{a\,\psi''(a)}{\psi'(a)}\right),
\end{multline*}
and the proposition is proved since last term is bounded away from
zero.
\end{proof}

\begin{example}\label{rem:beta0}
The above theorem applies to the power--law cases discussed in
Example \ref{ex:power}, with $\beta$ strictly positive, and more
generally for functions $\psi$ not--decreasing but strictly concave
in $(0,a_0)$.

The case $\beta=0$ (see also Remark \ref{rem:power1}), where
$|k^2\Psi|$ is seen to go to zero, may be treated using ideas from
\cite{nolan}, and results in a weak singularity.
\end{example}

\section{Discussion and conclusions}\label{sec:final}
As we have seen, the formation of singularities the gravitational
collapse of homogeneous scalar fields with potential is completely
ruled by a condition of integrability of a function related to the
energy density of the model. If this function is bounded, apparent
horizon formation is avoided during the evolution, and therefore
the singularity is massless, which, as is well known, is actually a feature of any
spherically symmetric naked singularity.

The boundary develops a naked singularity when the exterior solution
is given by anisotropic de Sitter generalization \eqref{eq:KS}, and
the singularity turns out to satisfy a strong curvature condition
for a class of examples that also includes power--law cases studied
in \cite{joshi}.

Let us briefly comment on the choice of the exterior region. In the
examples discussed in \cite{joshi}, the matching is performed with a
wider class, that is the generalization of Vaidya spacetime. One may
wonder whether the strong curvature condition is satisfied also
along null geodesics with this more general choice of mass profile.
Actually, in this case $k^2\Psi$ is not \emph{identically} zero --
as it happens in the case we studied -- but it can be checked that
again it vanishes \emph{in the limit} $k\to 0^+$, due to matching
condition \eqref{eq:addit}.

It must therefore be observed that, if the exterior region is given
by \eqref{eq:KS} or even the more general metric \eqref{eq:Va}, the
limit strong curvature condition proved in Proposition \ref{thm:tip}
does not hold for any radial geodesic, but only for timelike
geodesic satisfying  $\frac{\mathrm dY}{\mathrm dk}\to 0$ in the
approach to the singularity. Such a case is possible due to the fact
that no apparent horizon form in the interior region, and $\chi$ is
bounded away from zero near the singularity. Although one may try
and see what happens with other exterior metrics, this fact seems to
be a distinctive feature of the homogeneous scalar field model under
exam, which possess a synchronous singularity, and therefore is
qualitatively different from other Cosmic Censorship counterexamples
known in literature \cite{JoshiCMP,ns}.

In conclusion, breaking the homogeneity  assumption of scalar field
models with potential seems an unavoidable step to retain strong
curvature condition along \emph{any} causal geodesic terminating
into the naked singularity.

\end{document}